\documentclass [
  sigconf,
  screen,
] {acmart}

\usepackage[T1]{fontenc}
\usepackage[utf8]{inputenc}

\usepackage{
  cleveref,
  framed
}

\crefname{table}{Tbl.}{tables}
\Crefname{table}{Table}{Tables}
\crefname{figure}{Fig.}{figures}
\Crefname{figure}{Figure}{Figures}
\crefname{section}{Sect.}{sections}
\Crefname{section}{Section}{Sections}

\copyrightyear{2026}
\acmYear{2026}
\setcopyright{cc}
\setcctype{by}
\acmConference[FSE Companion '26]{34th ACM Joint European Software Engineering Conference and Symposium on the Foundations of Software Engineering}{July 05--09, 2026}{Montreal, QC, Canada}
\acmBooktitle{34th ACM Joint European Software Engineering Conference and Symposium on the Foundations of Software Engineering (FSE Companion '26), July 05--09, 2026, Montreal, QC, Canada}
\acmDOI{10.1145/3803437.3805238}
\acmISBN{979-8-4007-2636-1/2026/07}

\begin{document}

 \title
   [How to value open source contributions?]
   {How to Value Open Source Contributions?\\
    An Institutional Perspective from CERN}

\author{Julie Skoven Hinge}
\email{julie.skoven.hinge@cern.ch}
\orcid{0009-0002-7876-8602}

\affiliation{%
  \institution{IT University of Copenhagen}
  \country{Denmark}
}

\affiliation{%
  \institution{CERN}
  \country{Switzerland}
}

\author{Micha Moskovic}
\email{micha.moshe.moskovic@cern.ch}
\orcid{0000-0002-7638-5686}
\affiliation{%
  \institution{CERN}
  \country{Switzerland}
}

\author{Axel Naumann}
\email{axel.naumann@cern.ch}
\orcid{0000-0002-4725-0766}
\affiliation{%
  \institution{CERN}
  \country{Switzerland}
}

\author{Noemi Calace}
\email{noemi.calace@cern.ch}
\orcid{0000-0002-1494-9538}
\affiliation{%
  \institution{CERN}
  \country{Switzerland}
}

\author{Andrzej Wąsowski}
\email{wasowski@itu.dk}
\orcid{0000-0003-0532-2685}
\affiliation{%
  \institution{IT University of Copenhagen}
  \country{Denmark}
}

\begin{abstract}
We present a methodology to systematically assess the scale and impact of an organization's contributions to open source software (OSS). The methodology combines the archival data of Software Heritage with usage metrics, dependency analysis, economic valuation models, and interviews to comprehensively understand institutional OSS involvement. We then apply the methodology to the European Organisation for Nuclear Physics (CERN).  Despite using mostly commit data, we obtain a thorough overview of CERN’s OSS engagement. We identify over six million commits made to over 50,000 projects and highlight the most impactful projects led by CERN.\@  Beyond CERN, the methodology offers a reusable framework for organizations seeking to measure and evaluate their OSS contributions.
\looseness -1


\end{abstract}

\maketitle

\section{Introduction}

\emph {Open source software} (OSS) refers to software developed through open collaboration, enabled by licensing schemes that make its source code freely available for use, inspection, and modification. OSS has become a fundamental component of the modern software supply chain. The Linux Foundation estimates that 70-90\% of any modern software solution originates from OSS, making its continued development and maintenance essential to sustaining nearly all industries, both public and private~\cite{perlow2022censusii}.
One of the most prominent examples, the Linux Kernel itself, powers over 90\% of web servers and Internet-connected devices, serves as the foundation for Android, and runs on all of the world's top 100 supercomputers~\cite{whiting2022enterprises}.
\looseness -1

OSS is maintained not only by individuals in their free time, but also by companies and organizations that invest significant resources into its development. They donate software to the community, or allocate developers to contribute to existing OSS projects, often to influence the project's direction, integrate desired features, or to increase the company's reputation\,\cite{whiting2022enterprises,guizani2023rules}.
Institutions have been involved in contributing to OSS for decades\,\cite{lakhani2003hackers}. This long-standing ecosystem raises a central question for organizations: what is our impact, and how can we understand the scale and significance of our contributions?
Hoffmann and colleagues hypothesize a labour-market approach to estimate both the supply-side and demand-side value of widely used OSS. The supply-side cost is calculated as the replacement cost to recreate software once (i.e., what would it cost in labour to recreate the software from scratch), while the demand-side cost estimates how much firms would need to spend if OSS did not exist and they had to develop equivalent proprietary software. They find that the demand-side value of \$8.8 trillion significantly exceeds the supply-side cost of \$4.15 billion. This massive discrepancy suggests that policymakers and businesses should increase investment in OSS sustainability to maintain its economic benefits~\cite{hoffmann2024value}.
Similar work has also focused on measuring the popularity of OSS projects through GitHub usage metrics and download statistics, providing insights into their adoption and reach~\cite{dozmorov2018github,parekhexploring, openssf_criticality_score,yang2020makes,pfeiffer2021identifying}.
While these studies tend to focus on global and national trends, the global value of OSS, and the value of a select set of projects, we lack ways for an individual institution to understand its own contributions and impact in the OSS landscape. Moreover, most existing research relies on narrow data sources, commonly data scraped from GitHub, and therefore overlooks contributions hosted on other platforms such as GitLab or Bitbucket.
\looseness -1


This study is motivated by the above questions being asked inside the European Organization for Nuclear Physics (CERN).  Our goal is to design a methodology that allows for a comprehensive understanding of an organization's contributions to OSS.\@  In order to overcome the limitations of narrow data sources, we combine the comprehensive archival data of Software Heritage\,\cite{di2017software} with usage- and dependency-based valuation methods. Unlike the individual hosting platforms, Software Heritage provides a cross-platform graph that allows for a more comprehensive identification of institutional contributions. We ask two main research questions:%
\begin{itemize}

  \item[RQ1.] How can an institution comprehensively identify and understand its contributions to OSS projects?

  \item[RQ2.] How can value be assigned to these contributions, building on existing valuation models?

\end{itemize}%
RQ1 is answered through the use of metadata archived by Software Heritage, such as commits, authors, dates, and repository links~\cite{SH_graph}. We create a dataset that enables us to understand: (i) commit patterns in an organization's contributions, (ii) projects initiated in-house that attract external committers, and (iii) influential external projects to which the institution commits. To address RQ2, we assess value using several complementary indicators: repository-level popularity metrics on GitHub/GitLab (stars, forks, watchers), package-level usage statistics (downloads, dependency graph centrality), and estimated cost of recreation.
%
By combining contribution tracking through Software Heritage with usage-based valuation methods, this study aims to provide a data-driven assessment of an organization's involvement in the OSS ecosystem. In doing so, it not only sheds light on the scale and impact of open source commits stemming from one organization, but also contributes to a broader understanding of how institutional investments in OSS can be measured, valued, and sustained.

In the second part of the paper, we apply the method to CERN, painting a rich picture of CERN's importance in the open source landscape. We complement the quantitative analysis with interviews of CERN developers, showing how any institution can systematically measure and evaluate its role in the open source landscape.
\looseness -1

Our analyses show that CERN is a significant contributor to OSS, in diverse ways. We identified over 52,000 unique repositories associated with CERN, spanning over 6 million commits. Notably, many repositories are in areas not commonly associated with CERN's interests. These projects were often initiated by a single CERN contributor, either as part of their work or as a hobby project, and later attracted significant contribution and popularity by the developer community. Examples include \texttt{CLIUtils/CLI11}, \texttt{go-python/gopy}, and \texttt{vlachoudis/bCNC}. The analysis code and interview questions are available \cite{zenodo_code} to facilitate similar work for other organizations.

\section{Background and State of The Art}

\subsubsection*{Related Work}

Prior studies often focus on the economic value of OSS or on its popularity. For example Hoffman et al.\ estimate the economic value  from the demand and supply sides\,\cite{hoffmann2024value}. On the supply side, they measure the recreation cost of rewriting OSS from scratch. This is done using two datasets: (i) \emph{Census II of free and open source software (FOSS)---Application libraries}, a list of the most widely used OSS based on data from software composition analysis companies, and (ii) \emph{BuiltWith}, a dataset constructed by scanning websites to identify the underlying technologies. They use the COCOMO\,II model\,\cite{boehm2009software} to estimate the labour cost of rewriting OSS from scratch. They suggest that companies save trillions of dollars by using OSS.\@  Without, the industry would spend approximately 3.5 times more on software than they currently do.\@ Korkmaz and coauthors also use COCOMO\,II, but adjust costs to developer salaries by country and institution, based on contributor biographies on GitHub\,\cite{korkmaz2024github}. They group developers by country and sector, and the value of their contributions is calculated by multiplying the number of lines of code by the average national developer salary. They estimate that the US is the top investor in OSS, having contributed approximately \$37.8 billion in 2019.

Going beyond economic metrics, Yang et al.  define project impact as the ability to attract user attention\,\cite{yang2020makes}.  They find that the number of forks on GitHub is perceived as the strongest indicator of impact, followed by stars and watchers. 
In the bioinformatics domain, Dozmorov finds that GitHub stars, watchers, and forks have a modest correlation with the number of software citations for popular bioinformatics tools, and suggests that GitHub statistics should be used, in addition to traditional impact metrics, to measure the practical utility of bioinformatics software\,\cite{dozmorov2018github}.
While they highlight the connection between academic and GitHub recognition, in this work we are more interested in practical usage and community engagement as indicators of impact, rather than citations. In fact, research software is rarely cited directly, with on average only about 0.1 citations per software record\,\cite{park2019research}. The correlation between actual use of software and popularity indicators is also blurred. It turns out that a high star count on GitHub may indicate appeal within the developer community, but it does not necessarily reflect functional relevance (found by crossing GitHub and PyPI data\,\cite{parekhexploring}). Pfeiffer proposes an alternative measure of OSS importance by applying PageRank to dependency networks, highlighting projects that are highly connected and thus potentially critical to the ecosystem\,\cite{pfeiffer2021identifying}.
\looseness -1

OpenSSF has built a tool used for measuring the criticality of a project based on a number of signals, including the commit frequency, age of the project, the number of dependencies it has, and more. These signals are collected into a weighted score, assigning criticality to any project with a score between 0 and 1. They use this tool themselves to release a list of the top  586,751 most critical OSS as determined by their criticality score. This score aims to define the influence and importance of a project, and the dataset was released to make the OSS community especially aware of these projects to proactively improve their security~\cite{openssf_criticality_score}.

The above studies propose individual complementary metrics to capture the value of OSS.\@  Usage indicators, such as downloads, stars, and package dependencies, quantify adoption, while contribution metrics, such as the number of commits or diversity of committers, reflect engagement and activity. Economic measures, including the estimated cost of development or replacement, provide additional insight into the broader value of software. We want to use these metrics in combination to create a unified framework for assessing contributions to OSS from a single organization, and in doing so, aim to quantify both the scale and significance of these contributions.
\looseness -1

\subsubsection*{European Organization for Nuclear Physics}

CERN presents a rich and interesting case of an institutional contributor to OSS and open science, with a long-standing commitment to making research outputs publicly available and fostering global collaboration.
It was found\-ed  in the 1940s, as a world-class physics research facility in Europe for conducting research and facilitating international collaboration following World War\,II.\@ The CERN convention, signed in June 1953 by twelve member states, emphasizes that research conducted at CERN must be motivated by scientific use and that all results must be published or otherwise made public. Today, CERN has 25 member states who share the operating costs and oversee all decisions and activities by being represented in the Council\,\cite{cern_timeline_89}. In addition, CERN collaborates with numerous countries and organizations through cooperation agreements, scientific contracts, or other collaborations.
The main purpose of CERN is to conduct fundamental research in particle physics, and much of this research is being conducted through CERN's flagship accelerator, the Large Hadron Collider, a 27-kilometre accelerator that accelerates particles to nearly the speed of light, placed 100 meters underground\,\cite{cern_timeline_89}.
Beyond particle physics, CERN has made substantial technological contributions. For example, the World Wide Web was invented at CERN to help scientists collaborate on research\,\cite{cern_birth_web}. Other advancements include innovations in medical imaging\,\cite{cirilli2021_cern_medtech}, environmental applications\,\cite{cern_kt_environment}, aerospace\,\cite{cern_kt_aerospace} and computing\,\cite{cern_os_history}.
%
By the 1970s, computing had become critical for high-energy physics, which led CERN to invest in computing research such as the HBOOK plotting software, which was developed at CERN and shared with the Lawrence Berkeley National Laboratory in California and the Russian accelerator Laboratory at Serpukhov, who were close collaborators of CERN at the time\,\cite{cern_os_history}.
This culture of collaboration and openness resulted in revolutionary innovations such as the capacitive touch screen (1973) and the release of the World Wide Web into the public domain (1993), both innovations made public to the world with the belief that the sharing of these tools leads to a higher amount of innovation on a global scale, and thus will lead to collective progress\,\cite{cern_os_history}.
More recently, CERN released the research data management framework InvenioRDM, which powers platforms like Zenodo. In addition, CERN has created other widely used tools, including Indico, InspireHEP, and ROOT, all of which are released under open source licenses, further exemplifying its role in promoting open science and open source technology globally\,\cite{cern_legacy_open_science}.
\looseness -1

\section{Methodology} \label{sec:methodology}

We refer to the organization of interest, the contributing organization, as \emph{the Organization} from now on. Our method has two main steps: (i) identify OSS commits created-by and repositories contributed-to by the Organization and (ii) assess the impact of these commits within the broader ecosystem.
\looseness -1

\subsection{Identifying the Organizations' Commits}%
\label{sec:commits_extraction}

\begin{figure}[t]

  \includegraphics [scale=0.9] {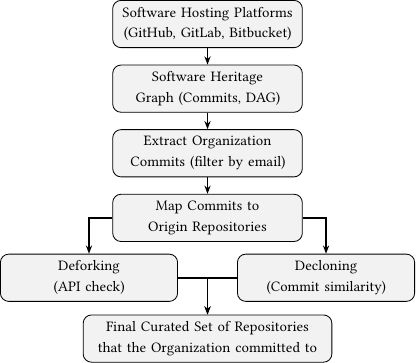}
  \Description{Flow diagram showing each step of the methodology from the Software Heritage graph, to extracting and mapping commits, to deforking and decloning, and finally obtaining the list of curated repositories.}
  \vspace {-0.8mm plus 0.2mm minus 0.2mm}

  \caption {Overview of the methodology for identifying and filtering Organization commits to OSS repositories.}%
  \label {fig:methodology}

  \vspace {-3.0mm plus 0.2mm minus 0.2mm}

\end{figure}

\begin{figure} [t]

  \includegraphics [
    width = 0.96\linewidth
  ] {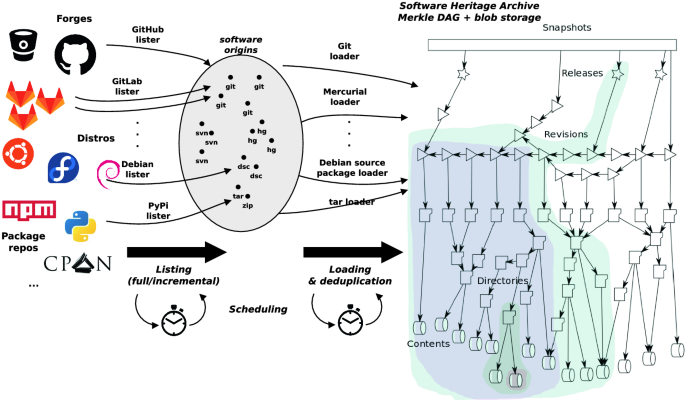}
  \Description{Diagram showing the process of gathering source code from hosting platforms and representing it in the Software Heritage Graph.}

  \vspace{-3.3mm plus 1mm minus 0.3mm}

  \caption{Gathering source code from hosting platforms and representing it in the Software Heritage Graph\,\cite{di2022should}.}%
  \label{fig:sh_process}

  \vspace {-3.0mm plus 0.2mm minus 0.2mm}

\end{figure}

\Cref{fig:methodology} summarizes our method for identifying Organization's commits and repositories.  While GitHub and GitLab provide hosting and collaboration mechanisms, Software Heritage is an \emph {archive} that \emph {aggregates} development data from many platforms. Software Heritage crawls platforms such as GitHub, Gitlab, or Bitbucket for software artifact information and stores it in a directed Merkle graph linking together source code files, repositories, and version control commits across time, capturing the global public software development landscape~\cite{softwareheritage}. Each artifact is uniquely identified.  Each commit includes metadata enabling detailed analysis.  The graph supports extraction of project-level metadata, including the origin (e.g., repository URL), release versions, and commit histories. \Cref{fig:sh_process} summarizes the Software Heritage collection process: data flow from diverse hosting platforms into a graph structure.

In this study, we use  “History and hosting” compressed Software Heritage graph, dated 24/12/06, and query it using a local gRPC server installation. The graph stores commits as \emph{revision nodes} referenced by Software Hash Identifiers (SWHID) of the form \texttt{swh:}\-\texttt{1:}\-\texttt{rev:\-<hash>}, where the last token is a Git commit hash\,\cite{iso18670}. We search the nodes to find the projects that the Organization has contributed to, and metadata from every commit authored by a member of the Organization is collected.  Each revision node contains:
\looseness -1

\begin{enumerate}
    \sloppy
    \item A pseudonymized author identifier (the person who originally wrote the changes) and a pseudonymized committer identifier (the person who actually recorded those changes into the repository; in many cases the same individual; they can differ, for example, in cases of code review).
    \item The commit message describing the change. 
    \item The commit timestamps of both the author timestamp (when the change was written) and the committer timestamp (when it was officially recorded in the repository).
\end{enumerate}

\noindent
To filter all commits made by the Organization, we match the email of both the author and committer entries against a list of selected pseudonymized addresses belonging to the employees from the Organization. By using both fields, we ensure that contributions made by affiliates of the Organization in either role are included.
It is important to note three limitations of this method when identifying commit authors belonging to the Organization:

\begin{enumerate}

    \item Not all Organization-affiliated developers use their work email when contributing to OSS projects, even when doing so as part of their work duties.

    \item If a developer changes their work email (e.g., to shorten the email), this person will appear as two unique authors.

    \item Some individuals with Organization-email addresses can be guest researchers, students, or external collaborators, rather than full-time staff.

\end{enumerate}

\noindent
We trace each commit in the graph to its origin SWHID in the original source repository, from which it was archived. This way we obtain a list of origin SWHIDs with at least one Organization-associated commit and the corresponding repositories.

Unlike approaches that rely solely on GitHub data, we leverage the comprehensiveness of the Software Heritage archive. As the largest public archive of source code, Software Heritage aggregates contributions from multiple hosting platforms. This ensures that commits made on GitLab, Bitbucket, or less prominent platforms are also captured, providing a complete and platform-independent view of the Organization’s engagement with the global OSS ecosystem.

\subsubsection{Organization's Commit Share}%
\label{sec:percentage}

We traverse all repositories (origin nodes) in the Software Heritage graph to identify all reachable commits (revision nodes). We compare the commits author's email against the list of predefined email addresses of the Organization. The commit share is calculated as the ratio of Organization-authored commits to the total number of commits in that repository.
\looseness -1

\subsubsection{Deforking}%

The above process collects also repository forks and clones, including many near- or complete duplicates that should be filtered out. We group repositories by the name seen in the URL, and query the platform API for each member in the group. We use the first found repository as the candidate for fork resolution. If the repository is itself a fork, the API is used recursively to extract its parent or source repository.  Once a candidate origin repository is identified, we cross-reference it against the initial dataset (above) to ensure that it has contributions from the Organization. Only repositories present in the initial dataset are retained in the final filtered set. Otherwise, the entire group of associated forks is discarded.
\looseness -1

\subsubsection{Decloning}

Developers sometimes duplicate an existing repository by downloading it or cloning it and re-uploading it under their own account without making an explicit fork in the hosting platform.  Such duplicates contain the original commit history, but are not removed by the above procedure, so they inflate the number of unique repositories committed to by the Organization.  We identify these duplicates by measuring the similarity of their commit histories. Specifically, for each repository, we use its history of commit timestamps extracted in \cref{sec:percentage} to compute the pairwise Jaccard similarity between these multisets, restricting the comparison to repositories with more than 50 recorded Organization commits in order to ensure statistical reliability. Pairs of repositories with a Jaccard similarity greater than 0.90 are considered duplicates. These pairs are then represented as edges in an undirected graph, where highly connected components represent clusters of repositories with highly similar commit history, indicating that one of these is the original repository, while the rest are copies. From each cluster, we retain only the repository with the highest degree centrality as the original repository, removing all others from the dataset.
\looseness -1

\subsection{Assessing OSS Impact} \label{sec:impact_methodology}

Measuring the impact of OSS projects is inherently challenging. Direct usage statistics are difficult to obtain due to the distributed and decentralized nature of OSS development.  We adopt a set of complementary impact metrics derived from prior works, capturing different dimensions of repository relevance and influence.

\subsubsection{GitHub/GitLab Statistics}

One way to gauge a project's popularity among developers is through its GitHub or GitLab stars, watchers, and forks.  Users star a repository to show appreciation (52\%), to bookmark it for later retrieval (51\%), or to signal interest in its use (36.7\%)\,\cite{gh_star}.
%
Watching a repository allows users to receive notifications about project's activity. Watching is associated with engagement, and 4.7\% of watchers later become committers\,\cite{sheoran2014understanding}. Some users discover interesting projects by studying the watching activity of influential users they follow\,\cite{dabbish2012social}.
Forking allows a developer to copy a repository, often to modify it or to submit contributions (46\%)\,\cite{jiang2017and}. 
%
Using the respective APIs, we collect the number of stars, watchers, and forks for each repository and rank repositories by the sum of these three metrics. As starring reflects appreciation, watching reflects engagement, and forking reflects contribution \cite{gh_star, sheoran2014understanding, dabbish2012social, jiang2017and}, these metrics are collectively a meaningful proxy for a repository's impact within the developer community.

\subsubsection {PyPI Statistics}

Python projects can be packaged and released on PyPI, with meta-data that allows to
reliably link source code repository with released package statistics. Specifically, the DaSEA dataset\,\cite{dasea} provides structured access to PyPI metadata, including package URLs (matching packages to repositories) as well as download counts\,\cite{pypistats} (a quantitative measure of adoption). Unfortunately, equivalent reliable datasets with URL mappings and download statistics are not readily available for other ecosystems\,\cite{dasea}. 
\looseness -1

We use DaSEA tools to recreate the most recently available package information (September 2025).  DaSEA enables us to (i) identify Organization-repositories that are distributed through PyPI, (ii) retrieve information about their versions, maintainers, and dependencies, and (iii) link packages to repositories via their URLs. Packages with many downstream dependents may be considered especially influential within the Python ecosystem, while those with high download counts indicate strong adoption by the user community. To complement these measures, we also apply the PageRank algorithm\,\cite{pagerank} to the PyPI dependency network. This highlights packages that lie at the central positions within the ecosystem by accounting not only for the number of dependents but also for the relative importance of those dependents. PageRank therefore provides a structural view of influence within the Python package ecosystem, beyond counts of downloads or direct dependencies.

The DaSEA data complements GitHub statistics with broader user-ecosystem-level indicators of popularity and influence, and captures both the availability of Organization projects on package managers and their integration into broader dependency networks.

\subsubsection{Replacement Cost} \label{sec:cocomo}

To estimate the dollar value of creating a piece of software from scratch, we first calculate the effort in person-hours using COCOMO\,II applied to all Organization repositories (following\,\cite{hoffmann2024value}). Formally, the effort for project $i$ is
\begin{equation}
E_i = \alpha \cdot \eta \cdot L_i^{\beta} \enspace,
\end{equation}
where $L_i$ is the project size in KLOC, $E_i$ is the effort in person-months, $\alpha=2.94$, $\beta=1.0997$ are default COCOMO\,II constants, and $\eta$ is the subjective adjustment (here $\eta = 1$). We use \texttt{pygount} to count lines of code. We convert the effort into a monetary value using a weighted global wage\,\cite{hoffmann2024value}. We identify the top 30 countries by active GitHub contributors in 2021 and collect the average annual salary of a software developer in each of these countries using Salary Expert. These salaries are converted to monthly and then hourly rates (assuming 40 hours per week and 4 weeks per month). Each country’s wage is weighted by its share of active contributors within the top 30 countries to produce a global hourly wage:
\begin{equation}
\textstyle W = \sum_{c=1}^{30} s_c \cdot w_c \enspace,
\end{equation}
where $w_c$ is the hourly wage in country $c$ and $s_c$ is the share of contributors in that country.
To focus on the contribution of the Organization itself, we further weight the estimated labor cost by the fraction of the total project contribution attributable to Organization developers. The total labor cost of project $i$ is therefore:
\begin{equation}
C_i = H_i \cdot W \cdot f_i \enspace,
\end{equation}
where $H_i$ is the total person-hours for project $i$ and $f_i$ is the fraction of contributions coming from Organization developers.

\subsubsection{Commits to Critical Projects} \label{sec:critical_methodology}

To identify impactful commits beyond Organization-originated repositories, we examine the Organization's involvement in the OpenSSF \textit{critical projects} dataset\,\cite{openssf_criticality}.
%
%
We match these projects against our list of repositories with Organization commits by comparing repository URLs. For each overlapping repository, we count the number of contributed commits.

To highlight the projects that are both highly critical and heavily committed to by the Organization, we designed a general-purpose metric: the \textit{Commit-Criticality Score (CCS)}. This score combines the number of commits from an organization with the OpenSSF Criticality Score of each repository. Both components are normalized to the $[0,1]$ range using min–max scaling, ensuring that each contributes comparably. Formally, for repository $i$:
\begin{equation}
\text{CCS}_i = \alpha \cdot \text{Criticality}_i^{\text{norm}} + (1 - \alpha) \cdot \text{Commits}_i^{\text{norm}} \enspace,
\end{equation}
where $\alpha$ is a weight (we use $\alpha = 0.5$). This way we balance projects that are extremely critical but have few commits with those that are moderately critical but where the Organization is very active.
\looseness -1

\section{Findings on CERN OSS Contributions}

We present the results of applying the method of \cref{sec:methodology} to identify and evaluate CERN’s contribution to OSS, providing a comprehensive view of CERN’s role in the global OSS ecosystem.

\subsection{Understanding CERN's OSS commits}


\begin{figure}[t]  

  \includegraphics [
    width = 0.96 \columnwidth,
    clip,
    trim = 6.7mm 16mm 7.1mm 15mm
  ] {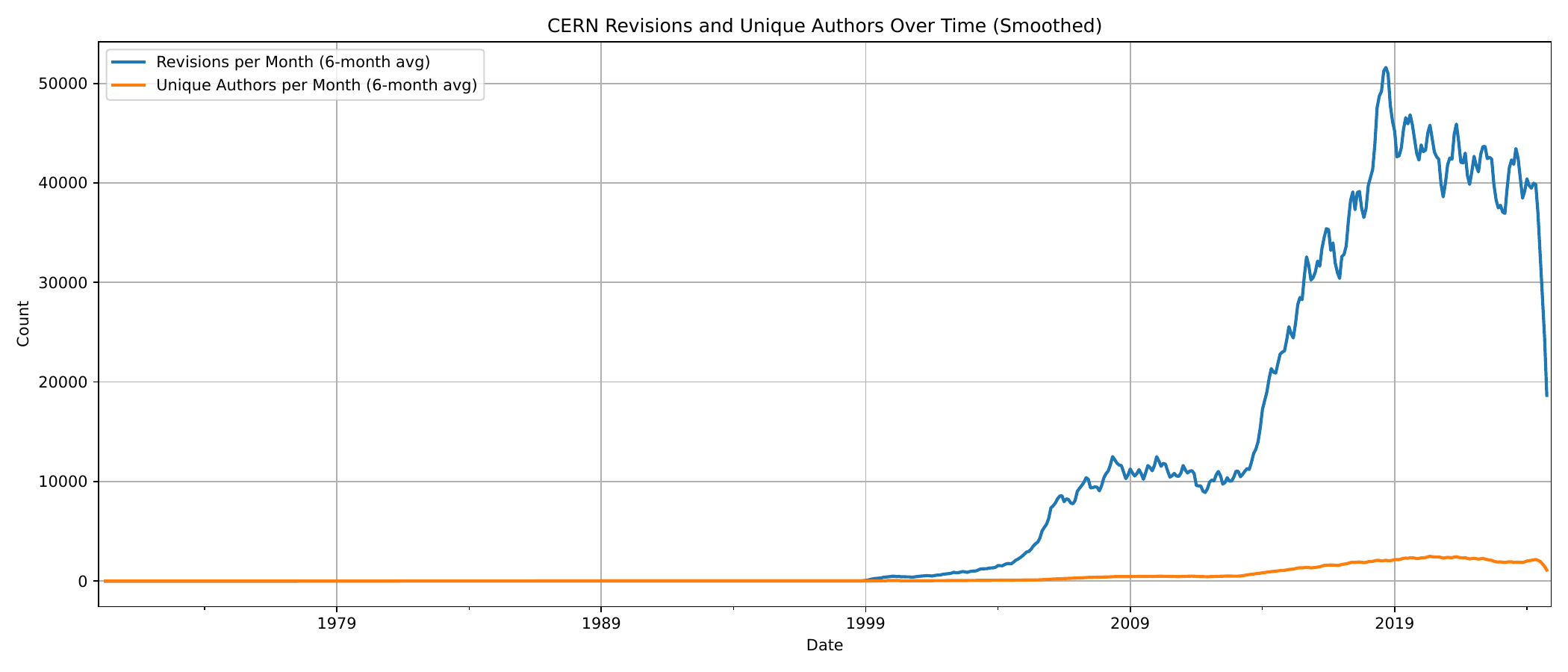}

  \vspace{-1.5mm plus 0.2mm minus 0.2mm}

  \caption{CERN commits and committers over time.}%
  \label{fig:over_time}

  \medskip

  \includegraphics [
    width = 0.96 \columnwidth,
    clip,
    trim = 3mm 2.5mm 2mm 11.5mm
  ] {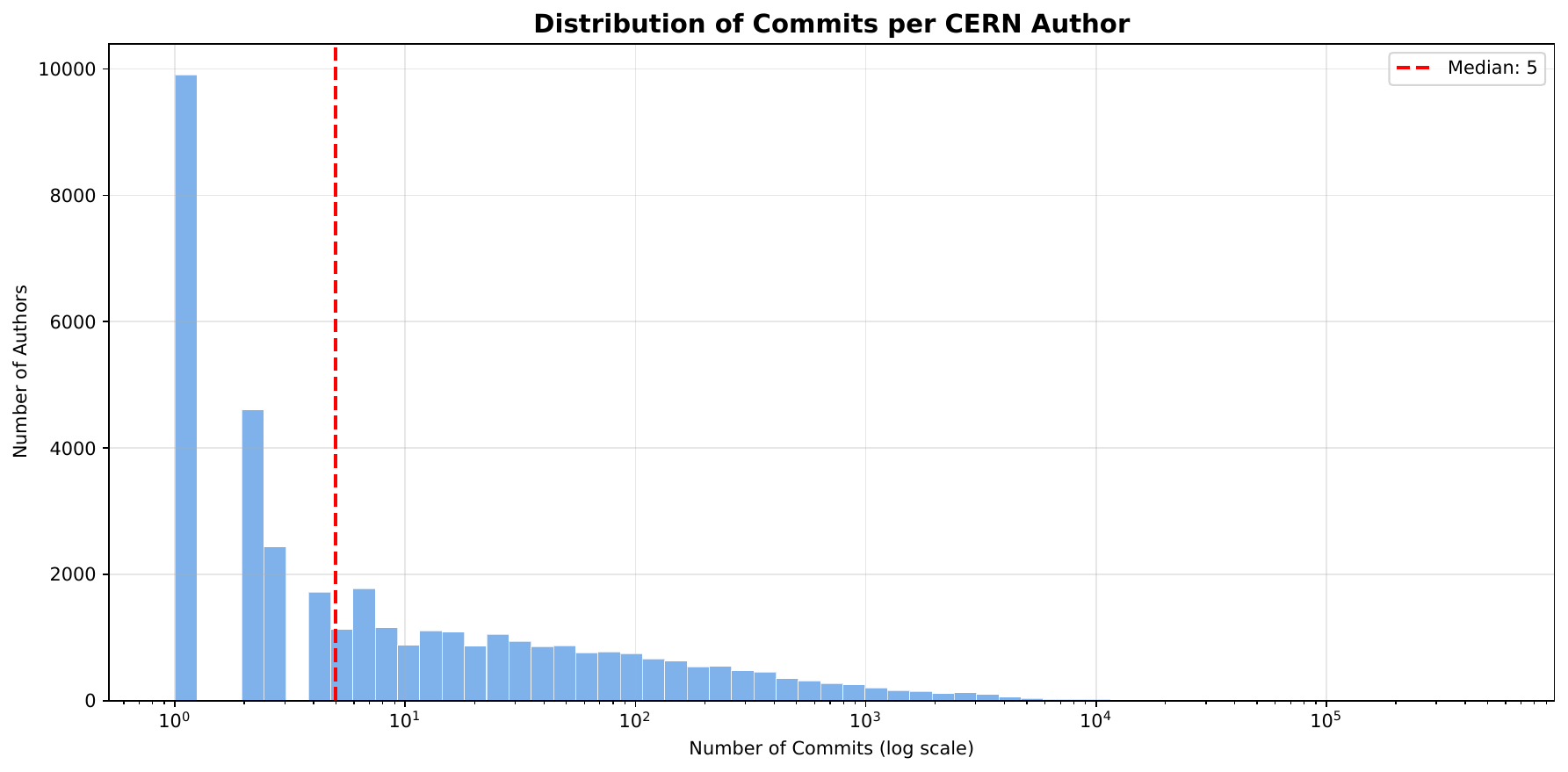}

  \vspace{-1.5mm plus 0.2mm minus 0.2mm}

  \caption{Commits per CERN-affiliated author.}%
  \label{fig:commit_histogram}

  \vspace{-2.5mm plus 0.2mm minus 0.2mm}

\end{figure}

\begin{table}[t]

  \small
  \begin{tabular}{@{}l r}
  \textbf{Metric} & \textbf{Value} \\
  \midrule
  Total commits (CERN authors) & 6,007,871 \\
  Unique CERN authors & 38,719 \\
  Public repositories with CERN contributions (excl. forks) & 52,166 \\
  CERN-associated repositories ($>$1\% CERN commits) & 39,799 \\
  \end{tabular}

  \medskip

  \caption{CERN commits to OSS, a quantitative overview}%
  \label{tab:basic_stats}

  \vspace{-3.0mm plus 0.2mm minus 0.2mm}

\end{table}

\begin{table*}[t]
\small
\begin{minipage}[b]{0.29\textwidth}
\centering
\begin{tabular}{lrr}
\textbf {Commit Range} & \textbf{Authors} & \textbf{Share} \\
\midrule
1                 & 9,906 & 25.9\% \\
2--10             & 13,276 & 34.8\% \\
11--100           & 9,262 & 24.2\% \\
101--1,000        & 4,704 & 12.3\% \\
1,001--10,000     & 992 & 2.6\% \\
10,001+           & 60 & 0.2\% \\
\\
\\
\\
\end{tabular}

\vspace{2mm}
\caption{Authors by commit range}%
\label{tab:author_distribution}
\end{minipage}%
\hfill
\begin{minipage}[b]{.29\textwidth}
  \centering
\begin{tabular}{lr}
\textbf{Percentile} & \textbf{Contributions} \\
\midrule
25th   & 1 \\
50th   & 5 \\
75th   & 36 \\
90th   & 201 \\
95th   & 504 \\
99th   & 2,697 \\
99.9th & 12,862 \\
\\
\\
\end{tabular}
\vspace{2mm}
\caption{Commit/author percentiles}%
\label{tab:contrib_percentiles}
\end{minipage}%
\hfill
\begin{minipage}[b]{.40\textwidth}
\centering
\begin{tabular}{l r}
\textbf{Language} & \textbf{Percentage of repositories} \\
\midrule
Python            & 27.5\% \\
C                 & 22.6\% \\
C++               & 18.9\% \\
Unknown           & 8.8\% \\
HTML              & 6.3\% \\
Jupyter Notebook  & 5.9\% \\
Ruby              & 3.8\% \\
Shell             & 3.7\% \\
JavaScript        & 2.4\% \\
\end{tabular}
\vspace{2mm}
\caption{Main programming language of repositories with CERN contributions}%
\label{tab:language_percentages}
\end{minipage}

\vspace{-4.5mm plus 0.2mm minus 0.2mm}

\end{table*}

\subsubsection{Dataset Overview}

We followed the method of \cref{sec:methodology} and identified 6,007,871 CERN-contributed commits in 1,100,728  repositories. Filtering for forks and clones resulted in 52,166 unique CERN-associated repositories, whereas 39,799 of those have a CERN commit percentage of over 1 percent (cf.\ \cref{tab:basic_stats}). It is hard to find reliable data on this.  However, Alphabet, widely seen as a large contributor to OSS, informally reports \emph{interactions} with 70k+ (presumably \emph{non-unique})  repositories in 2023 \cite{Vargas_2024}. In the same period, CERN has \emph{contributed} to 6011 \emph {unique} repositories.

\Cref{fig:over_time} aggregates the number of CERN commits and contributing authors over time. The oldest (found) commit by a CERN affiliate was made in late 1993. All commits spanning from September 1993 until April 1996 were made towards the W3C Protocol Library, part of the World Wide Web infrastructure. The most recent commits, in this study, were made at the end of 2024 (determined by the version of Software Heritage graph we use). A peak is visible in 2018.  Within this peak, we identify the following example repositories with increased activity relative to their other active years: \texttt{trackml-library} (utilities for the trackML challenge\,\cite{LAL_trackml_library}), \texttt{go-zeromq/zmq4} (a pure-Go implementation of 0MQ\,\cite{go-zeromq_zmq4}), and \texttt{scikit-hep/root\_pandas} (a Python module for loading/saving ROOT files as DataFrames\,\cite{scikit-hep_root_pandas}).

A total of 38,719 unique authors with a CERN email address had made at least one commit. \Cref{fig:commit_histogram} shows the distribution of commits per author. The distribution is highly skewed, with the majority of authors contributing little, and a long tail of very prolific committers. The median number of commits per author is five. Out of the top ten committers, three are automated accounts (bots), including the one that made the highest number of commits. According to an informal report from Yahoo!Finance, large contributing organizations have between two and six thousand \emph{active} contributors.\footnote{\url{https://fosslife.org/5-biggest-open-source-contributors.html}, 2022 data, seen 2026/1}  Thus it, again, appears that CERN is a significant OSS contributor.
\looseness -1

\Cref{tab:author_distribution} summarizes the distribution of authors across commit ranges, while Table~\ref{tab:contrib_percentiles} reports key percentiles. More than 60\% of authors have made ten or fewer commits, while only 2.8\% have made more than 1,000. The top 1\% of authors (382 individuals, each with at least 2,697 commits) account for 50.7\% of all commits.
\begin{leftbar}
  \noindent
  Even in this large organization, most of the commits are made by a small number of individuals.
\end{leftbar}

\noindent
\Cref{tab:language_percentages} lists the most common programming languages by repository that CERN commits to (computed with GitHub/GitLab API).

\begin{figure}[t]  
  \includegraphics [
    width = 1.015\columnwidth,
    clip,
    trim = 2.8mm 9mm 6.2mm 2.8mm
  ]{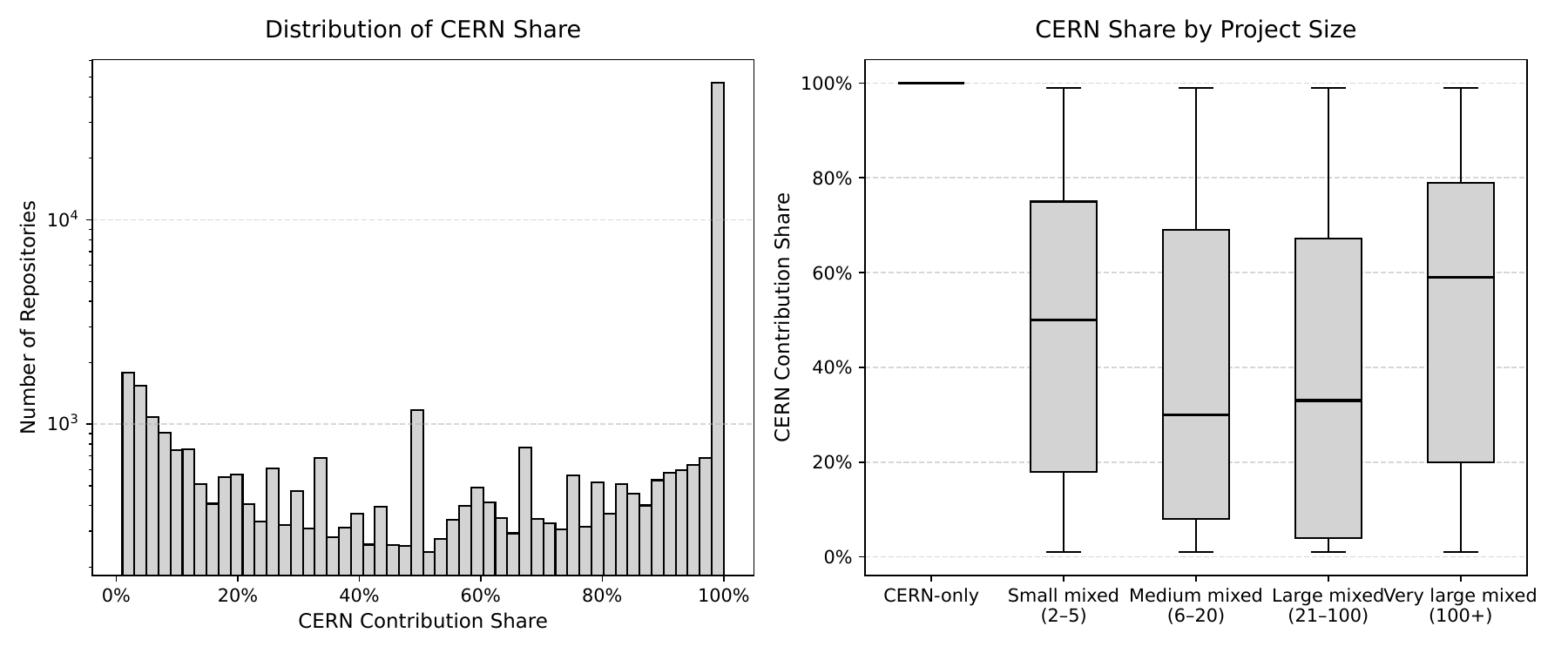}
  \Description{Plot of CERN commits share per repository.}

  \vspace{-4.0mm plus 0.2mm minus 0.2mm}

  \caption{CERN share per repository}%
  \label{fig:cern_share_combined}

  \vspace{-4.0mm plus 0.2mm minus 0.2mm}

\end{figure}

\subsubsection{Commit Share Across Repositories}

\Cref{fig:cern_share_combined} summarizes the commit patterns across the 52,166 repositories. The left panel shows the overall histogram of CERN commit shares.  The right panel illustrates how CERN's share varies across project sizes (measured by number of unique committers). The mean repository has 80\% CERN commits, and 68\% of repositories are at least 90\% CERN-driven. Most repositories are majority CERN-led: 80.4\% have $\geq 50\%$ CERN contributions, while only 4.6\% fall between 1--5\%.
This shows that while CERN is a minority contributor in some external collaborations, the majority of repositories in the dataset are institution-led projects where CERN provides most of the development.
Examples of such include \textit{ROOT}, \textit{Invenio}, and \textit{Zenodo}, while examples of projects where CERN has a contribution percentage of less than 1 percent include \textit{Tensorflow}, \textit{curl}, and \textit{Kubernetes}.

\begin{leftbar}
  \noindent
  When assessing organizational contributions to OSS it is useful to distinguish owned projects from wider collaborative efforts.
\end{leftbar}

\begin{framed}

\noindent\textbf{RQ1: How can an institution comprehensively identify and understand its contributions to OSS projects?} We show that an institution’s OSS contributions can be systematically identified by leveraging the cross-platform archival data of Software Heritage. By filtering commits using institutional identifiers and mapping them to origin repositories, it is possible to recover both institution-led projects and contributions to external OSS infrastructure at scale. Unlike GitHub-centric approaches, this method captures contributions across multiple hosting platforms and enables analyses of contribution volume, author distributions, and institutional commit shares across repositories. While commit-based identification does not capture all forms of participation, it provides a scalable and reproducible foundation for institutional OSS analysis.

\end{framed}

\vspace{-2mm plus 0.2mm minus 0.2mm}

\subsection{Impact of CERN Repositories}


\subsubsection{GitHub/GitLab Statistics}%
\label{sec:github_stats}

\begin{figure*}[t]

  \includegraphics[
    width = 0.7\textwidth,
    clip,
    trim = 2.3mm 9mm 2mm 15.3mm
  ] {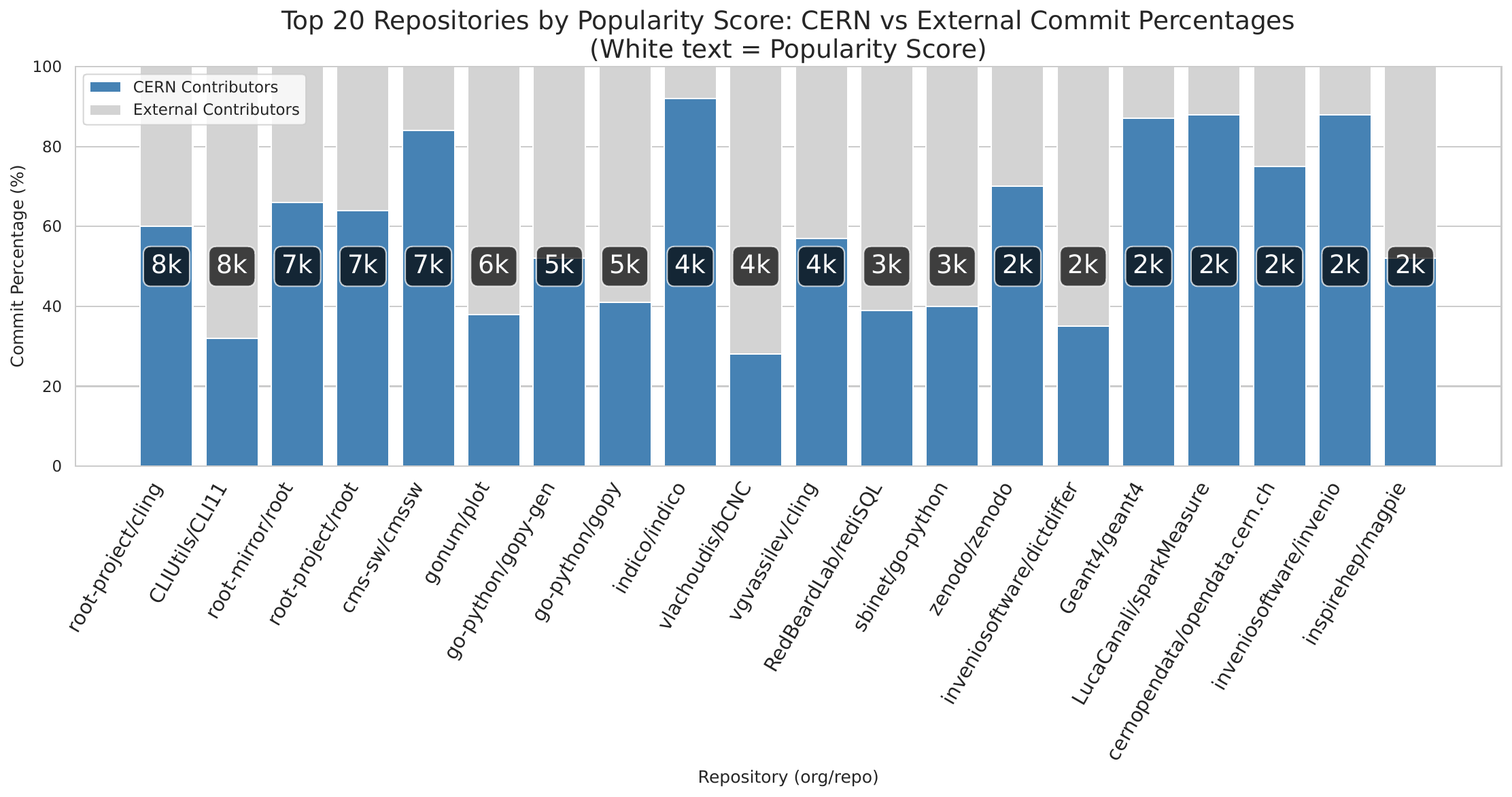}
\Description{A bar plot showing the contribution percentage share between CERN and external developers of the top 20 repositories ranked by popularity score.}
  \vspace{-3.5mm plus 0.2mm minus 0.2mm}

  \caption{Top twenty CERN repositories by GitHub popularity score (stars + forks + watchers, the score is in white)}%
  \label{fig:pop_score}

\end{figure*}

GitHub stats are available for over 91.2 \% of CERN repositories. For this analysis, we focus on repositories with over 20\% commits from CERN  (26,849) with a combined number of 72,235 stars, 58,672 forks, and 52,326 watchers. We rank these repositories by the amount of combined stars, watchers, and forks earned on GitHub/GitLab. \Cref{fig:pop_score} lists the top 20 repositories.
\looseness -1

Out of these repositories, 12 were started at CERN. \texttt{Cmssw} is software developed specifically for one of CERN's largest experiments, and most outside contributions likely come from collaborators or employees not using their CERN email. Meanwhile, \texttt{cling}, \texttt{root-mirror/root}, and \texttt{root-project/root} are all part of the ROOT software ecosystem for storage, processing, and analysis of scientific data. Several projects are physics-related, while others serve more general scientific computing purposes. \texttt{Zenodo} and \texttt{Indico} are tools used for research output repositories and event management, respectively. Interestingly, five repositories have only one CERN committer yet maintain a fairly high CERN commit percentage, even though there are many external committers as well (\texttt{plot}, \texttt{gopy-gen}, \texttt{gopy}, \texttt{vlachoudis/bCNC}, and \texttt{go-python}).

\begin{figure}[t]
\centering

  \includegraphics [
    width = 0.75 \columnwidth,
    clip,
    trim  = 0mm 0mm 0mm 3mm
  ]{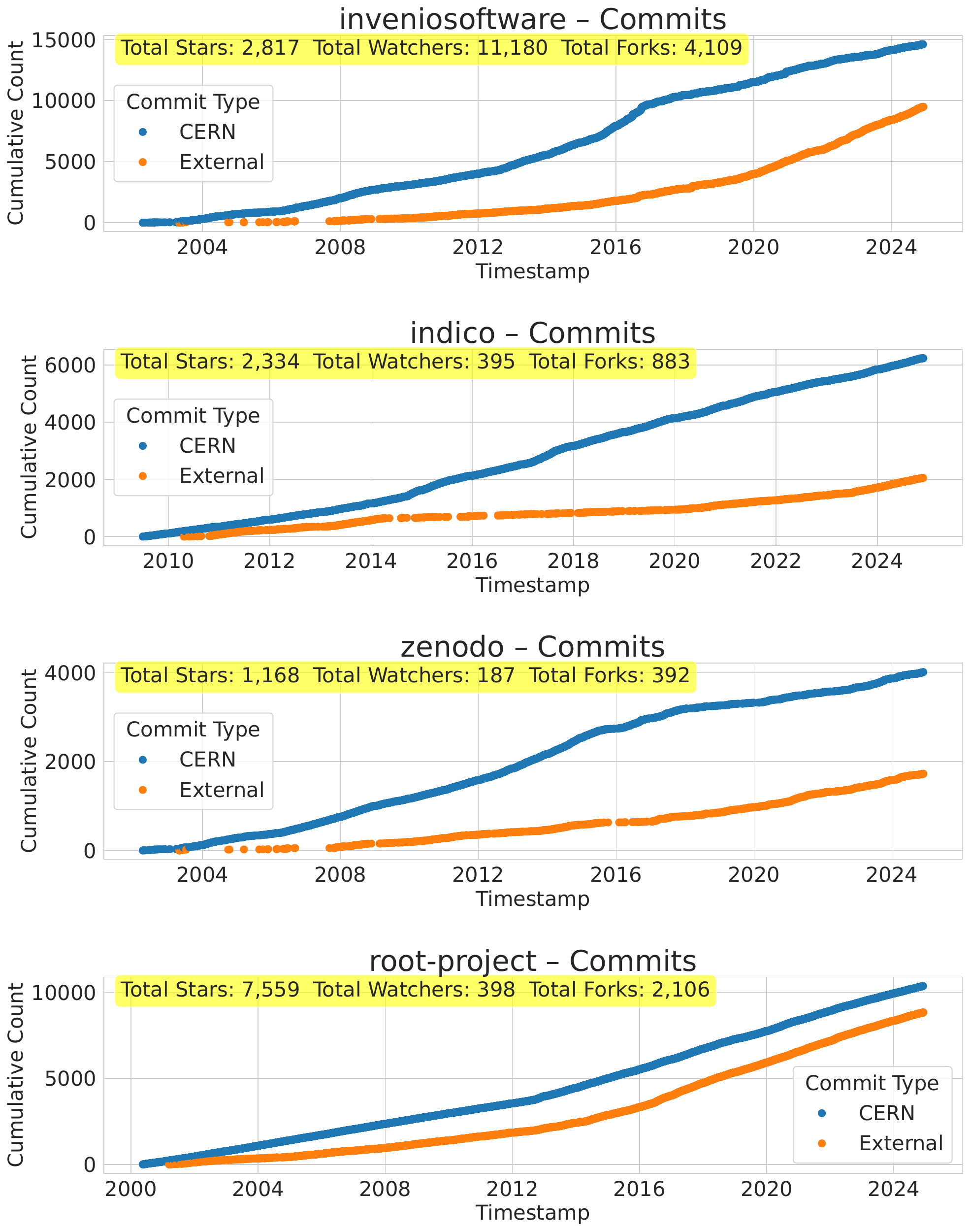}

  \vspace{-3mm plus 0.2mm minus 0.2mm}

  \caption{CERN vs external development over time}%
  \label{fig:temporal}

  \vspace{-3.5mm plus 0.3mm minus 0.3mm}

\end{figure}

To gain further insight, we explore the temporal commit patterns of four selected GitHub projects originating from CERN (\cref{fig:temporal}). We separate internal and external contributions, and note that although CERN dominates the commit share, all four projects have attracted a growing external contribution.
\looseness -1

\begin{figure*} [t]

  \centering
  \includegraphics [
    width = .7 \textwidth,
    clip,
    trim = 0mm 9mm 0mm 15mm
  ] {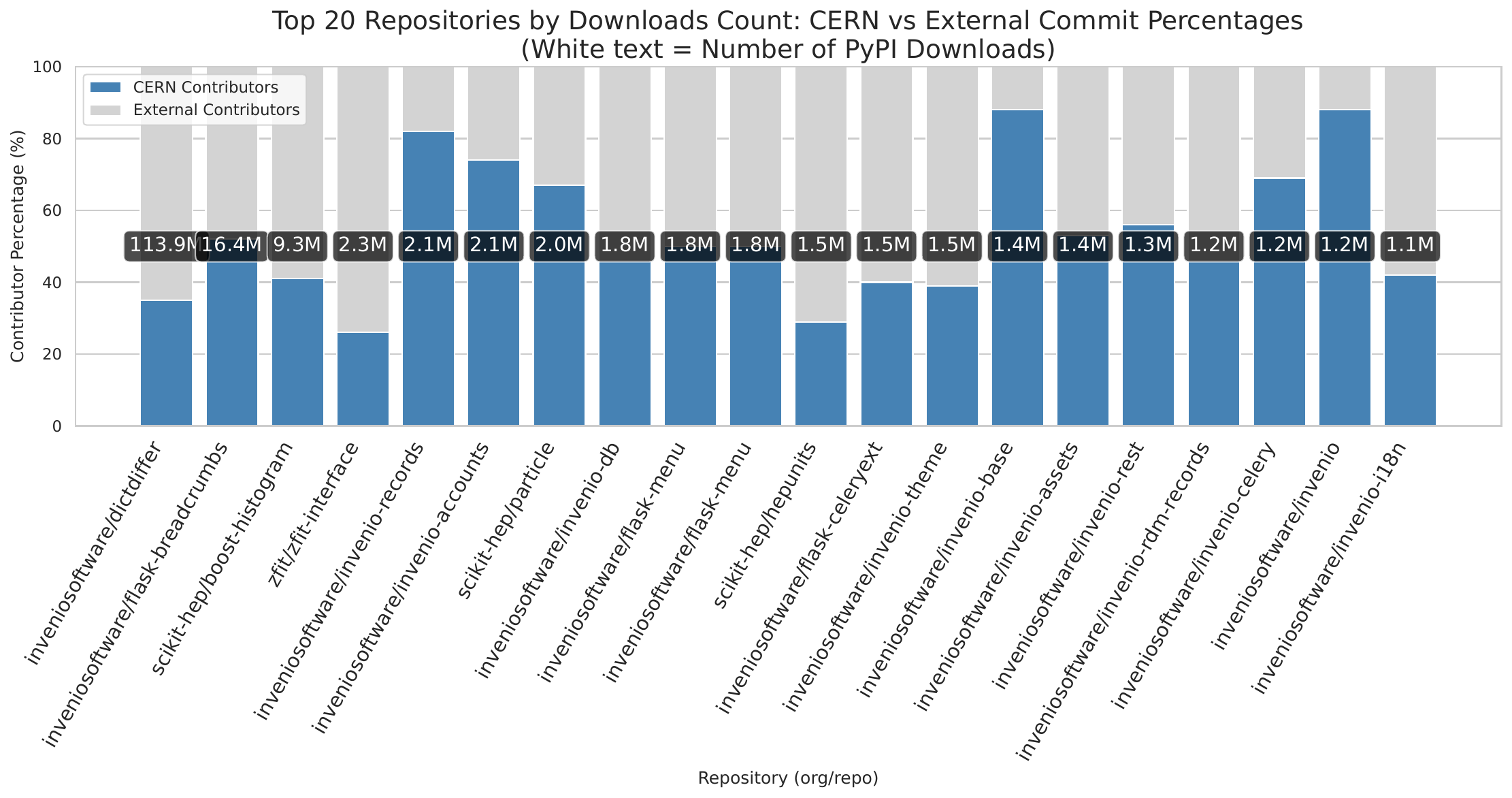}

  \vspace {-3mm plus 0.3mm minus 0.3mm}

  \caption{Top twenty CERN repositories ranked by number of PyPI downloads (in white)}%
  \label{fig:downloads_score}

  \vspace {-3mm plus 0.3mm minus 0.3mm}

\end{figure*}

\subsubsection{PyPI Statistics}

Out of 26,849 repositories with over 20\% CERN commits, 352 are distributed on PyPI, with 208,228,793 PyPI downloads in total.
We showcase the top 20 repositories ranked by the number of downloads in \cref{fig:downloads_score}. The packages related to Invenio, originally developed and maintained at CERN, are among the most highly downloaded. The prominent example is \texttt{dictdiffer}, a module designed to compute and apply differences between dictionaries. More broadly, Invenio serves as a versatile software framework that can be used as a repository management platform, an integrated library system, and a code library for building large-scale information systems. The \texttt{dictdiffer} package has 844 GitHub stars and, according to GitHub’s dependency graph, is used in over 29.5k other projects. We also find strong interest in packages related to scikit-hep, a Python ecosystem for High Energy Physics data analysis, many of which were initiated at CERN and continue to receive significant contributions from CERN developers. Applying the PageRank algorithm to the PyPI dependency network further highlights these results. \texttt{dictdiffer} ranks 1880th globally within the entire PyPI network, with 167 direct dependents. Indico, CERN’s event management platform, appears at rank 2171, with 31 dependents.
\looseness -1

\begin{table}[t]

  \small
  \begin{tabular} {
    l
    r
    @{\hspace{1mm}}r
    @{\hspace{1mm}}r
  }

  \textbf{Repository}
  & \llap{\textbf{CERN Contributions}}
  & \textbf{Criticality}
  & \textbf{CCS}
  \\

  \midrule
  \texttt{cms-sw/cmssw}                 & 353515 & 0.72023 & 0.9111 \\
  \texttt{jcmvbkbc/linux-xtensa}        & 60     & 0.84704 & 0.5001 \\
  \texttt{root-project/root}            & 104498 & 0.61746 & 0.4869 \\
  \texttt{torvalds/linux}               & 60     & 0.80952 & 0.4738 \\
  \texttt{ceph/ceph-client}             & 61     & 0.80936 & 0.4737 \\
  \texttt{gregkh/linux}                 & 63     & 0.80003 & 0.4671 \\
  \texttt{hisilicon/linux-hisi}         & 60     & 0.79764 & 0.4655 \\
  \rlap{\texttt{multipath-tcp/mptcp\_net-next}} & 60    & 0.79646 & 0.4646 \\
  \texttt{Xilinx/linux-xlnx}            & 61     & 0.78971 & 0.4599 \\
  \texttt{linux-audit/audit-kernel}     & 62     & 0.78749 & 0.4584 \\
  \end{tabular}

  \bigskip

  \caption{The most critical projects with CERN contributions, ranked by Commit-Criticality Score (CCS)}%
  \label{tab:critical_projects_weighted}

  \vspace{-4.5mm plus 0.3mm minus 0.3mm}

\end{table}

\subsubsection{Contributions to Critical OSS Projects}

Following \cref{sec:critical_methodology}, we find that CERN developers contributed to 1,555 of the 586,751 most critical open source projects in the OpenSSF dataset. The commits are not concentrated on a single repository, but are spread over a wide range of projects. CERN involvement is not confined to CERN-maintained software only, but it extends into widely used projects across scientific and infrastructure domains.
On the other hand, as many as 103 critical projects have over 20\% of CERN contributions, so they are developed intensively at CERN. Examples include domain-specific software such as \texttt{root}, and infrastructures like \texttt{indico} and the Invenio ecosystem.  \Cref{tab:critical_projects_weighted} lists the top 10 repositories ranked by CCS, which balances criticality and CERN contributions. Note that CERN contributes to critical repositories in a small scale, and that some significant CERN-developed projects are ranked moderately on the criticality score.
\looseness -1

\begin{framed}

\noindent\textbf{RQ2: How can value be assigned to institutional OSS contributions?}
We show that the value of institutional OSS contributions cannot be captured by a single metric but can be approximated through a combination of complementary indicators. Usage-based measures such as downloads and dependency counts are signals of adoption, while repository-level metrics offer limited insight into visibility and community interest. To contextualize institutional engagement in widely used OSS, we introduce the Commit-Criticality Score (CCS), which combines an organization’s volume of commits with project-level criticality. When complemented with qualitative insights from developer interviews, this multi-dimensional approach enables a more nuanced understanding of both the scale and relevance of institutional OSS contributions.

\end{framed}

\section{Deep-dive into Selected Projects}

To further understand the collaboration dynamics and impact of CERN's OSS activities, we conduct a qualitative analysis of four representative, but different, projects: \texttt{Indico}, an event management system for meetings and conferences, \texttt{Zenodo}, an online repository for research outputs, \texttt{InvenioRDM}, a research data management platform that Zenodo was built on, and \textit{ROOT}, an analysis framework for high energy physics data. We carry out semi-structured interviews with developers involved in these projects\,\cite{zenodo_code}.
We selected these projects as they consistently rank among the most impactful CERN repositories (Figures~\ref{fig:pop_score} and \ref{fig:downloads_score}) and have established, accessible development teams at CERN.
\looseness -1


\subsubsection*{Indico}

Indico is a platform where users can schedule meetings, lectures, conferences, and store all related material. First a joint initiative between CERN and four European research institutions (2022), it was taken over by CERN after two years\,\cite{ferreira2015indico}.  Indico has ca.\ 400,000 users, including  many besides CERN, e.g., Max Planck, UN, and UNESCO\,\cite{edge2025indico}. The Indico GitHub organization has 2,334 stars, 395 watchers, and 883 forks. Following \cref{sec:cocomo}, we estimate the replacement cost of Indico to be US\$3,330,502 with global weighted salaries and \$2,971,524 when scaled by the CERN commit share (\$7,157,690 and \$6,386,198 respectively if Swiss salaries were used).
\looseness -1

Out of the 57,194 Indico revisions, 50,452 were contributed by CERN, and 6,742 came from external contributors. Out of 300 unique authors, 83 used a CERN email, 95 used a Gmail account (responsible for 2,207 revisions), 7 had a UN email address (869 revisions), and 49 used GitHub-generated emails (778 revisions, some from automated accounts such as Dependabot). Three authors with addresses from unconventional.dev, a consultancy founded to develop Indico extensions, and three from Fermilab (113 revisions). Other contributions came from a diverse range of institutions and domains, including universities such as the Technical University of Munich and Cornell University, as well as private individuals and IT companies.
\looseness -1

The two interviewed developers emphasized that while most external contributions are feature-driven, such as the conditional registration fields introduced by the UN Indico development team, they often take other forms. For example: \emph{``the other cool thing, that actually the UN contributed as well, is [that] they hired a guy to work on accessibility improvements for Indico, which is really cool… all of our stuff is now slowly becoming more and more compliant.''} Similarly, translations into more than fifteen languages are provided almost entirely by volunteers (\emph{“all the other translations that we have in Indico are done by volunteers for free from the community”}).
CERN’s role in the project is less about co-developing features with external actors and more about integrating and reviewing external work. \emph{“It’s not super collaborative in the sense that they come to you and you’re like, oh, okay, we also want this. Now we’re going to work on it together. It’s more that they work on it themselves and then you review it and decide whether you want to integrate it or not”}. Many external contributions are one-off fixes from individuals, though some return consistently, particularly in areas like translations.
\looseness -1

Funding is also a contribution model: after the initial European project ended in 2004, external support has come primarily through targeted grants or contracts, such as those from the Sloan Foundation or the Max Planck Institute. As one interviewee puts it: \emph{“they either give us money or they hire somebody to work on it… they contribute much less, (...) targeted features that they want.”}

Despite the widespread adoption, the team does not track detailed usage metrics, citing privacy considerations. \emph{“We don’t track that at all for privacy reasons. The only thing we track is [if] the instance [is] alive, is it responding, and the version.”} There are ca.\
300 active installations with opt-in tracking, at research organizations like Fermilab and INFN, at universities, charities, and small companies.
\looseness -1

Interestingly, the developers define success not in terms of commits, downloads, or user counts, but in terms of community recognition: \emph{“people just come to us randomly and say like, hey, thanks for working with Indico. It’s really cool. We love it.”}


\begin{leftbar}
  \noindent
  Feedback through conference interactions, surveys, or direct expressions of gratitude, are considered more meaningful indicators of impact than numeric measures.
\end{leftbar}

\subsubsection*{Zenodo/InvenioRDM}

We analyzed both \texttt{Zenodo} and its underlying framework \texttt{InvenioRDM}, to understand contributions at the application and platform levels. Zenodo is a general-purpose repository for research outputs ranging from datasets and software to preprints and reports. Zenodo was launched by CERN and OpenAire in 2013. Predominantly developed within CERN, it has attracted external contributions over time\,\cite{zenodo10years}. Zenodo is the EU-recommended repository for research results, but anyone in the world can freely deposit information in the platform hosted on CERN’s computing infrastructure. By its 10th anniversary, Zenodo reached over 300,000 users across 7,500+ institutions in 153 countries, hosting more than 1 PB of data and attracting tens of millions of views and downloads annually\,\cite{zenodo10years}. It is used intensively also in the Software Engineering research community. Zenodo was initially developed from the Invenio software framework, which has evolved into three main projects by today: InvenioRDM, InvenioILS, and Invenio Framework\,\cite{zenodo10years}.
\looseness -1

The Zenodo organization has 1,168 stars, 187 watchers, and 392 forks, while InvenioRDM has 2,817 stars, 11,180 watchers, and 4,109 forks. The estimated replacement cost of Zenodo is \$1,208,251 when using global weighted salaries, and \$819,963 when scaled by the CERN commit share (\$2,596,692 and \$1,762,209 respectively at Swiss level). For the Invenio software framework, the replacement cost is \$44,417,517 in global salaries, and \$24,660,747 scaled to the CERN share (\$95,459,115 and \$52,999,206, respectively at Swiss level).

Out of 4,069 commits in the Zenodo organization, 2,531 are contributed by CERN, while 1,538 come from outside. We found\,117 unique authors: 23 from CERN and 94 external. Many externals use Gmail accounts (803 commits) or GitHub-generated emails (240), but some represent institutions such as getlektor.com, linux.com, uni-konstanz.de, zhbluzern.ch, and simmons.edu (between 8 and 346 commits each). While CERN contributors are fewer, they tend to be more active per author than the external ones.

In comparison, the underlying framework, InvenioRDM, shows a broader community of contributors. We analysed 66,607 commits, of which 39,376 originated from CERN emails and 27,231 from outside. There were 170 unique CERN and 656 external committers. Among the external contributors, several institutions are particularly active: while contributors using Gmail made 15,254 commits, we see 1,500 commits from tugraz.at, 999 from tuwien.ac.at, 424 from kth.se, 245 from caltech.edu, and 103 from cfa.harvard.edu.  Other contributions came from smaller organizations and personal accounts worldwide. On average, CERN committers made 222 commits each (median 19), while external committers made 42 commits each (median 3).

\begin{leftbar}
  \noindent
  This shows a highly active participation of CERN developers alongside a vibrant globally distributed external community.
\end{leftbar}

\noindent
Historically, Invenio laid the foundation as CERN’s repository software, powering the CERN Document Server, while Zenodo was developed to \emph{host} diverse research outputs.
As one developer explains, \emph{“we told people… just don’t fork, just don’t copy Zenodo because it’s not meant to be copied. It’s just meant to be a service. And then people, despite us telling them not to do it, they still went ahead and did it.”}
This dynamic led to the creation of InvenioRDM, designed as a framework for research data management. It allows institutions to build their own repositories, while benefiting from CERN's development investment. As one interviewee puts it: \emph{“one of the big decisions we took was, okay, we do InvenioRDM, we move CDS [the CERN Document Server] and Zenodo into InvenioRDM and build the same platform… because at some point [as] Zenodo is externally funded, ... the minute the funding is not there, it’s a problem.”}
Consequently, Zenodo and InvenioRDM go hand in hand: Zenodo allows sharing open data and other research outputs at scale, while InvenioRDM provides developers with reusable infrastructure to create repositories.
\looseness -1
\begin{leftbar}
\noindent
Unlike Invenio, released in 2002, which struggled to attract sustained contributions, InvenioRDM was explicitly designed to integrate external partners into its development.  Today, external maintainers \emph{outnumber} CERN maintainers. This is a fundamental transformation from the limited CERN-centered community around Invenio to a collaborative development model.
\looseness -1
\end{leftbar}
\noindent
An interviewee summarizes: \emph{“we have two main assets. One is InvenioRDM itself, the platform, the other thing is a welcoming community. And it’s the community that you are part of something that makes people contribute and that they feel that they are heard, that they are seen, that their needs [are taken into account].”}

The projects were shaped by EU funding, collaborations, and community involvement. Zenodo, born from the OpenAIRE initiative, grew into a trusted service due to CERN's reputation.
Zenodo and InvenioRDM rely heavily on a distributed team of partners. The core team at CERN remains small (4--7 people). The majority of maintainers are external. \emph{“Most of the maintainers are external at the moment… between eight to ten people. Most of them are the partners that started the project in 2016–15 and they’ve been there since the beginning.”}
The interviewees emphasise that external contributions take many forms. Partners provide community managers, run sprints, or contribute infrastructure. \emph{“Contribution is not just code… the project code is the least of the fifth. It’s all the reasons why do we do this? What do we want to build? Why did we come together?”} This reflects a broad understanding of value, where successful contributions include translations, outreach, testing, and even commercialization of services based on Invenio.
The declared motivations are linked to both professional and ideological factors: \emph{“People only contribute if they want to. It’s a personal thing, that’s for sure. Zenodo is one of the most important factors… people want to contribute to this open science movement and be part of it and make things better.”}
When asked about measuring adoption and success, the developers admitted that no comprehensive telemetry exists. Instead, they rely on indirect indicators such as known installations, community engagement, and external recognition. \emph{“We don’t have any telemetry that tells us there are 20 installations active or 100. So we don’t know that”} one developer admits. Yet Zenodo’s societal impact is evident: with around 500,000 users, it has become \emph{“the world’s largest general-purpose repository.”} A recent study estimated its economic impact at around 100 million Swiss francs per year\,\cite{crespo2024value}, but developers still highlighted the symbolic weight of recognition: \emph{“the only thing that works for upper management is that people from outside come and say, thank you for Zenodo.”} Another interviewee says: \emph{“ a community that is generally interested in the project, and they want to use it, they want to run it—that’s success.”}
\begin{leftbar}
\noindent
Success, in this sense, is defined not just by usage statistics or monetary estimates but by the sustainability of the community, the trust, even the happiness, it builds.
\end{leftbar}

\subsubsection*{ROOT}

ROOT was initially created in 1994 to address the complex data modelling needs of high-energy physics experiments, specifically the Large Hadron Collider.  It has evolved significantly over time, and not much of the original implementation remains after multiple major rewrites.  
ROOT is developed for high-energy physics but is also widely used beyond CERN in fields such as nuclear physics, gravitational wave research, finance, genomics, and aviation safety.
It boasts 7559 stars, 398 watchers, and 2106 forks. We estimate the replacement cost of the ROOT repository as \$2,102,666 at global salary level and \$1,408,575 when scaled by the CERN commit share (\$4,518,907 and \$3,027,213 respectively at Swiss level).
\looseness -1

The ROOT GitHub organization hosts 192,265 commits; 114,249 originate from CERN, while 78,016 commits come from outside. We found 245 unique CERN committers and 710 unique externals. Here, CERN developers had a median of 12 commits (average 466), while external committers had a median of 4 commits (average 110). Notable external contributions include 28k commits from GSI Helmholtz Centre for Heavy Ion Research, 16k from Fermilab, 1.3k from UCSD, 1.2k from National Institute for Subatomic Physics (Nikhef), and 995 from Lawrence Berkeley National Laboratory.

Interviews with the ROOT project lead highlight that its development has always been collaborative and international. While CERN is the largest contributor, ROOT receives many contributions from outside.
Presently, the CERN team consists of about 10 developers. Since the external contributors are highly dynamic, it is difficult to quantify how many there are, but nonetheless, they are considered crucial for the project's sustainability. The external contributions stem from institutional collaborations and from individuals: \emph{``People are interested and see value in contributing back...not only fixes or improving the stuff they are interested in, but they also felt it was right and nice to do something more.''} As with the Indico and Zenodo interview, the interviewee reflected that contributions are not only limited to commits, but that bug reports, reproducible test cases, and community engagement are also highly valued: \emph{``Commits is one aspect. clearly a bug report with a nice reproducer is also extremely precious...the project code is the least of it.''}
When asked about the success metrics of ROOT, the interviewee emphasizes practical impact over usage metrics, especially considering that usage is extremely difficult to track with OSS. Indicators of success include how effectively the framework supports the Large Hydron Collider programmes, the degree of external contributions, and the liveliness and engagement of the broader community: \emph{``One metric for success is how much people from outside (...) contribute...is the project alive? Do people care about contributing back?''}


\vspace{-3mm plus 0.4mm minus 0.2mm}

\section{Limitations and Validity}

\subsubsection*{Commit-based analyses.}

Identifying institutional contributions via email addresses is under-approximating, as developers may use personal accounts. Furthermore, commit-based analyses capture code-centric contributions and exclude non-code activities such as documentation, issue tracking, community outreach, and governance work. Insights from interviews with project leads do confirm that these non-code activities are central to OSS sustainability, yet they remain largely invisible to repository-based quantitative methods.
\looseness -1

Furthermore, while commits are a convenient and widely used unit for quantifying contributions at scale, they are an imperfect proxy for effort or impact. Commits vary substantially in size and significance, ranging from minor bug fixes to large architectural changes. As a result, figures and analyses that rely on commit counts may overemphasize frequent small contributions while under-representing fewer but more substantial changes. Alternative measures, such as the number of lines of code added or modified, offer complementary perspectives, but they introduce their own limitations, including sensitivity to coding style and refactoring practices. In particular, external contributors may disproportionately appear as low-impact contributors when assessed via commit counts, despite their changes addressing critical bugs or maintenance tasks. These factors should be considered when interpreting commit-based comparisons across projects and contributor groups. For a large contributor, like CERN,  however these problems are somewhat mitigated by scale---the law of large numbers implies that the variation is averaged across individual preferences and project policies. The proposed Commit-Criticality Score helps further to contextualize the contribution volume with project-level criticality, rather than relying on absolute commit counts alone.
\looseness -1

\subsubsection*{The data foundation.}

Software Heritage archives publicly accessible repositories, excluding private institutional development. Moreover, handling forks introduces trade-offs: while filtering forks helps reduce duplicate counting of shared commits, it may also remove projects that originated as forks but later evolved independently. All these under-approximate the scala of institutional contributions.
\looseness -1

\subsubsection*{Assessing impact and value.}

Usage-based indicators (e.g.\ download counts) can be incomplete or inflated. Repository-level metrics often reflect popularity rather than practical utility. Similarly, economic replacement-cost models, such as COCOMO II, give only coarse approximations and fail to capture factors like quality, long-term maintenance cost, or non-code contributions. Despite these limitations, we believe that our method is already a useful first step to systematically assess institutional contributions at scale.
\looseness -1

\section{Concluding Remarks}

This study shows how institutional involvement in OSS can be analyzed at scale by combining archival source code data with usage- and dependency-based indicators. Using the archive of Software Heritage, we mapped CERN's OSS activity beyond GitHub, capturing contributions across hosting platforms, and observing not only the scale of the engagement but also the diversity of its roles, as a producer of institution-led projects (ROOT, Zenodo, Invenio, Indico) and as a contributor to widely used shared infrastructure (operating systems, container orchestration platforms). Distinguishing these roles highlights how organizations benefit \emph{and} contribute through participation in shared technological foundations.
\looseness -1

The framework is modular, adaptable to organizations of different sizes by focusing on subsets of repositories, languages, or value indicators depending on available resources. For institutions like CERN, the analyses can inform strategic decisions regarding OSS investment, maintenance prioritization, and policy development. More broadly, this framework provides a foundation for future studies of OSS engagement across academic, industrial, or public-sector organizations, supporting policy making and funding decisions.
\looseness -1

\looseness -1

\begin{acks}
  This work was made possible by Software Heritage, the universal source code archive: \url {https://www.softwareheritage.org}. We thank Helge Pfeiffer for help with the DaSEA tool, Dawn Foster for suggestions on assessing impact, and CERN OSPO for input and feedback.
\end{acks}

\balance
\bibliographystyle{ACM-Reference-Format}
\bibliography{references}

\end{document}